\documentclass[runningheads]{llncs}
\usepackage[T1]{fontenc}
\usepackage{graphicx}
\usepackage{amsmath}
\usepackage{amsfonts}
\usepackage{booktabs}
\usepackage[table]{xcolor}
\usepackage[dvipsnames]{xcolor}
\usepackage{arydshln}
\usepackage[colorlinks=true,allcolors=RoyalBlue]{hyperref}

\usepackage[T1]{fontenc}
\usepackage{lmodern}
\usepackage[most]{tcolorbox}
\usepackage{enumitem}

\definecolor{promptframe}{HTML}{2F3E52}
\definecolor{promptbg}{HTML}{F7F9FC}
\definecolor{tagcolor}{HTML}{0B6E4F}
\definecolor{thinkcolor}{HTML}{8A5A00}
\definecolor{rulecolor}{HTML}{C9D3E0}

\newcommand{\ptag}[1]{\textcolor{tagcolor}{\texttt{<#1>}}}
\newcommand{\thinktag}{\textcolor{thinkcolor}{\texttt{<think>}\dots\texttt{</think>}}}
\newcommand{\lit}[1]{\texttt{#1}}

\newcommand{\phead}[1]{\par\smallskip{\bfseries\sffamily\color{promptframe}#1}\par\nobreak\smallskip}

\newtcolorbox{promptbox}[1][]{%
  enhanced, breakable,
  colback=promptbg,
  colframe=promptframe,
  boxrule=0.9pt,
  arc=3pt,
  left=6pt, right=6pt, top=4pt, bottom=4pt,
  fonttitle=\bfseries\sffamily,
  coltitle=white,
  attach boxed title to top left={xshift=10pt, yshift=-2mm},
  boxed title style={colback=promptframe, arc=2pt, boxrule=0pt,
                     left=6pt, right=6pt, top=2pt, bottom=2pt},
  #1
}

\begin{document}
\title{Learning Multi-Step Query Rewriting via Corpus Feedback for Conversational Search}
\titlerunning{Multi-Step Query Rewriting via Corpus Feedback}

\author{João Coelho\inst{1}\thanks{Work done while interning at Amazon.} \and
Hong Wang\inst{2} \and
Jie Yuan\inst{2} \and
Zhuoer Wang\inst{2} \and
Samson Koelle\inst{2} \and
Wei Niu\inst{2}}
\authorrunning{Coelho et al.}
\institute{INESC-ID and Carnegie Mellon University \and Amazon \\ \email{jmcoelho@andrew.cmu.edu}} 
\maketitle              
\begin{abstract}
Conversational Query Rewriting (CQR) turns a context dependent user turn into a standalone query for a retriever, and most methods do this in a single step from the dialogue history before retrieving once. The rewrite is therefore fixed before any corpus evidence is available to correct its reference resolution or its vocabulary. We recast CQR as a sequential retrieval problem: an agent rewrites the current turn, retrieves, and conditions its next rewrite on the returned passages. The agent acts in a typed space of three rewriting operations, resolving conversational intent into a standalone query, generating lexical reformulations, or synthesizing pseudo-documents for document-to-document matching, together with a stop action that ends the episode. We train the policy with supervised fine-tuning followed by reinforcement learning against a single retrieval-quality reward, using no human rewrite annotations. Across TopiOCQA and QReCC, the agent outperforms several retrieval-aligned baselines, while remaining effective across retrieval backends and generalizing to the CAsT benchmarks without additional training. Further analysis shows that, through retrieval-reward optimization alone, the learned policy develops a behavior of grounding pseudo-documents in passages retrieved by earlier steps, substantially improving retrieval.

\end{abstract}

\section{Introduction}\label{sec:intro}

Conversational search systems must handle queries whose meaning depends on prior turns~\cite{mo2024convsearchsurvey,zamani2023cis}. A question such as ``when was it released?'' cannot be resolved without the entities and topics established earlier in the conversation. Conversational Query Rewriting (CQR) transforms the current turn into a self-contained query that an existing retriever can process, and it is the standard interface between conversational systems and retrieval infrastructure~\cite{anantha2021qrecc,elgohary2019canard,vakulenko2021qr,yu2020fewshot}.

Most CQR methods produce a single rewrite from the conversational history and issue it directly to the retriever. This includes both supervised rewriters trained on human annotations~\cite{lin2020t5qr,vakulenko2021qr,yu2020fewshot}, and recent reinforcement learning methods that optimize rewriting from retrieval rewards. ConvSearch-R1~\cite{zhu2025convsearchr1}, for example, learns to rewrite using retrieval feedback but still makes a single rewriting decision before seeing any retrieved documents. Consequently, these methods cannot adapt the query to the language and entities used in the target corpus. Using retrieval results to inform a subsequent query is, however, a long-standing idea in information retrieval.  Pseudo-relevance feedback methods use terms from top-ranked documents to expand the initial query for a subsequent retrieval~\cite{abduljaleel2004rm3,rocchio1971,lavrenko2001relevance}. This principle has also been applied to conversational search, where retrieved documents are used to expand or refine conversational queries~\cite{lin2021hqe,park2024guidecqr}. Such approaches typically rely on predefined feedback mechanisms, fixing how information from retrieved documents is incorporated into the subsequent query rather than learning this process from retrieval outcomes.

Recent agentic search systems extend this iterative paradigm by alternating between search and reasoning until enough evidence is gathered to answer a question~\cite{jin2025searchr1,trivedi2023ircot}. They are typically trained and evaluated for end-to-end question answering, where retrieval is an intermediate step. How a conversational query should be reformulated after observing retrieved evidence, and how successive retrievals can progressively refine it, remains comparatively unexplored.

In this work, we formulate conversational query rewriting as a sequential retrieval problem. An agent repeatedly reformulates the current turn, issues each reformulation to the retriever, and uses the returned passages as feedback for its next decision. It chooses among three typed rewriting operations: (i) resolving conversational intent into a self-contained query, (ii) generating lexical reformulations, and (iii) synthesizing pseudo-documents for document-to-document matching~\cite{gao2023hyde}; it can also issue a stop action that ends the episode. We train the policy with supervised fine-tuning followed by reinforcement learning, using retrieval quality as the only objective and no human rewrite annotations.

We first evaluate results in the standard in-domain setting, training and evaluating on TopiOCQA and QReCC with BM25, and comparing against existing retrieval-aligned CQR methods under the same setup. We then test generalization across retrieval backends, by replacing BM25 with a dense retriever at inference time. Finally, we test zero-shot transfer by evaluating the TopiOCQA- and QReCC-trained models on TREC CAsT without further training. Across these settings, the multi-step agent improves over multiple baselines, with further analysis showing that retrieval-reward optimization alone leads the learned policy to progressively ground its reformulations in corpus evidence: it first resolves the query intent, optionally refines its lexical formulation, and finally generates pseudo-documents conditioned on passages retrieved by earlier steps.
\section{Related Work}

\paragraph{Conversational Query Rewriting.} Conversational search requires resolving the context dependence of user turns before retrieval~\cite{mo2024convsearchsurvey,zamani2023cis}. One line of work trains conversational dense retrievers that encode the full dialogue history into a query representation~\cite{lin2021cqe,mao2024chatretriever,mao2022cotec,yu2021convdr}, which couples context resolution to a retriever that must itself be trained. Conversational query rewriting instead produces a standalone query that any off-the-shelf retriever can consume, which is the setting we adopt. Early rewriters learned from human rewrites collected in datasets such as CANARD, QReCC, and CAsT~\cite{anantha2021qrecc,dalton2019cast,elgohary2019canard}, either as sequence-to-sequence generators~\cite{lin2020t5qr,vakulenko2021qr,yu2020fewshot} or as term selectors~\cite{voskarides2020quretec}. Since human rewrites are written for readability rather than retrieval, subsequent work aligns the rewriter with the retriever through retrieval-based rewards~\cite{wu2022conqrr}, preference- or ranking-based optimization over candidate rewrites~\cite{lai2025adacqr,yoon2025retpo,zhang2024adaqr}, retrieval-guided iterative data refinement~\cite{jang2024itercqr}, or by augmenting the rewrite with generated answers~\cite{mo2023convgqr} or LLM-enhanced conversation history~\cite{mo2024chiq}.

With LLMs, prompted rewriting became competitive without fine-tuning the rewriter~\cite{lai2025adarewriter,mao2023llm4cs,ye2023informative}, and ConvSearch-R1~\cite{zhu2025convsearchr1} removes external rewrite supervision by optimizing an LLM rewriter with reinforcement learning against a rank-shaped retrieval reward. ICR~\cite{cao2025icr} departs from single-step generation by alternating self-posed clarification questions with progressively refined rewrites, learned from retrieval-filtered trajectories through supervised and preference training, and fusing the rankings of all intermediate rewrites. Its refinement, however, operates on the query and dialogue alone: no step observes retrieved passages. Across these methods, retrieval signals, when used, enter only at training time, while the rewrites issued at inference are produced without observing the corpus. We retain annotation-free, retrieval-aligned training, but condition each rewrite on the passages returned by the previous ones.

\paragraph{Pseudo-Relevance Feedback.} Using the top of an initial ranking to refine a subsequent query is a long-standing idea in IR, from Rocchio's vector-space update~\cite{rocchio1971} to relevance models such as RM3~\cite{abduljaleel2004rm3,lavrenko2001relevance}, and has been carried over to dense retrieval, either by training a query encoder on feedback passages~\cite{yu2021anceprf} or by expanding the query with embeddings drawn from them~\cite{wang2023colbertprf}. LLMs shifted expansion towards generation, synthesizing pseudo-documents or answers that are matched against the collection~\cite{gao2023hyde,jagerman2023qe,mackie2023grf,wang2023query2doc}. Because such text is produced from parametric knowledge, it can diverge from the facts and vocabulary of the target corpus, which motivates conditioning generation on retrieved passages~\cite{lei2024csqe,shen2024lamer}.

Feedback has likewise been applied to conversational search, expanding the current turn with history terms selected by their behavior on the collection~\cite{lin2021hqe} or with keywords and expected answers extracted from initially retrieved passages~\cite{park2024guidecqr}. In these approaches the feedback mechanism is fixed a priori, i.e., the number of rounds, what is extracted from the passages, and how it is merged into the query are all hand-designed. We instead propose that an agent learns from retrieval reward which reformulation to apply at each step and when to stop. Pseudo-document generation is one of the agent's actions, grounded in passages retrieved at earlier steps rather than generated zero-shot.

\paragraph{Reinforcement Learning for Search Agents.} 
Reinforcement learning has been used to train query rewriters against retrieval or answer rewards~\cite{jiang2025deepretrieval,ma2023rewrite,mao2024rafe}, typically emitting a single rewrite per input. In parallel, agentic search systems interleave reasoning with repeated calls to a search engine, first through prompting~\cite{li2025searcho1,press2023selfask,trivedi2023ircot,yao2023react} and more recently through reinforcement learning, e.g., with GRPO~\cite{coelho2026effectivereinforcementlearningagentic,jin2025searchr1,shao2024deepseekmath,song2025r1searcher,zheng2025deepresearcher}. These agents are trained for end-to-end question answering, where retrieval is an intermediate step. Even approaches that train the searcher separately from the generator reward it through downstream answer quality~\cite{jiang2025s3}, so query reformulation is rarely studied on its own.

ChatR1~\cite{lupart2026chatr1} and ConvAgent~\cite{mo2026convagent} bring this paradigm to conversational question answering, interleaving search and reasoning within each turn and densifying the answer reward with intermediate signals, namely intent-aware turn-level rewards on search queries and the coverage of the gold answer by retrieved passages, respectively, the latter alongside a reward for mixed-initiative actions. Both remain optimized primarily for answer quality and require reference answers for training. We differ in objective and action space. Our policy is rewarded solely on the ranking induced by the final reformulation, requiring only passage relevance labels, with neither answers nor rewrite annotations. It also acts through a typed set of reformulation operations, which lets us analyze how reformulation strategies are sequenced in response to corpus feedback.

\section{Methodology}

This section formalizes the task (\S\ref{sec:formulation}), defines the agent and its typed action space (\S\ref{sec:design}), and describes the the training pipeline (\S\ref{sec:training}).

\subsection{Task Formulation}\label{sec:formulation}

A conversational search session is a sequence of user queries and system answers. At turn $t$, the user issues a query $q_t$ that may contain ellipsis or topic shifts relative to the conversation history $C_{t-1} = \{(q_1, a_1), \ldots, (q_{t-1}, a_{t-1})\}$. The retrieval objective is to rank the relevant passages for $q_t$ as high as possible in a collection $\mathcal{D}$, using a retriever $\mathcal{R}(\cdot)$ that scores passages against a query string.

Single-shot CQR instantiates this as learning a function $r_t = f(q_t, C_{t-1})$ and issuing $r_t$ to $\mathcal{R}$ once. We instead formulate the problem as an episode: given $(q_t, C_{t-1})$, an agent interacts with $\mathcal{R}$ over multiple steps, issuing intermediate rewrites and observing their retrieval results, before committing to a final retrieval state. The episode is internal to resolving turn $t$, and it is distinct from the conversational turns themselves.

\subsection{Agent Design}\label{sec:design}

At each step $s$ of the episode, the agent observes the conversational history, the current query, and the retrieval results of all previous steps. Then, it issues one of the following actions:

\begin{itemize}
\item \texttt{<intent>}: produce a self-contained reformulation of $q_t$, resolving references and ellipsis against $C_{t-1}$. This action targets the gap between the surface query and the underlying information need. 

\item \texttt{<rephrase>}: produce a lexical variant of the current formulation, targeting vocabulary mismatch between the query and the relevant passages.

\item \texttt{<hyde>}: produce a pseudo-document that may answer the resolved query, enabling document-to-document matching~\cite{gao2023hyde}.

\item \texttt{<stop>}: terminate the episode.
\end{itemize}

Each of the first three actions emits a tag followed by a rewrite, which is issued to $\mathcal{R}$, and the top-$k$ passages are appended to the context. Each subsequent action is therefore conditioned on corpus feedback, i.e. the evidence retrieved by earlier rewrites, rather than on the conversation alone. The episode ends when the agent emits \texttt{<stop>} or reaches a step budget of $T$ steps. The actions differ in how much they can exploit corpus evidence. While intent resolution can operate on the conversation alone, rephrasing and pseudo-document generation can draw on the vocabulary observed in retrieved passages. The system prompt specifies no order among the actions, and the agent may issue any action at any step.

\subsection{Training}\label{sec:training}

We train the agent in two stages, namely using a supervised fine-tuning stage that instills the action format, followed by a reinforcement learning stage that optimizes rewrite quality against the retriever.

\paragraph{Supervised Fine-Tuning.} Following common practice in RL-based agent training~\cite{jin2026behaviorprime,zhu2025convsearchr1}, we precede reinforcement learning with a supervised fine-tuning stage that addresses a cold-start problem where the small base model that has never emitted the action format produces mostly invalid trajectories, leaving GRPO without gradient signal. Since we train small models, we distill trajectories from a larger model of the same family running the agent harness described in Section \ref{sec:design}, and retain trajectories that follow the action format. The resulting checkpoint reliably emits typed actions but is not yet aligned with the retriever; that alignment is the role of the RL stage. 

\paragraph{Reinforcement Learning.} We train the SFT checkpoint with GRPO~\cite{shao2024deepseekmath}. Given a conversational turn instance $x = (q_t, C_{t-1})$, the behavior policy produces a group of $K$ trajectories $G(x) = \{o_1, \ldots, o_K\}$, where each trajectory is a full episode of interleaved actions, rewrites, and retrieval results. Each trajectory receives a scalar reward $r_i = R(o_i, x)$ described below. We compute group-normalized advantages from the rewards:

\begin{equation}
A_i = \frac{r_i - \mu_G}{\sigma_G + \epsilon_{\text{norm}}},
\qquad
\mu_G = \frac{1}{K}\sum_{j=1}^{K} r_j \;,
\end{equation}

\noindent where $\sigma_G$ is the standard deviation of rewards in the group and $\epsilon_{\text{norm}}$ is a small numerical constant. The trajectory-level advantage $A_i$ is assigned to every valid token in $o_i$, with tokens belonging to retrieved passages being masked from the loss. The policy is optimized with the clipped objective:

\begin{equation}
\mathcal{J}(\theta) = \mathbb{E}_{x, G}\!\left[
\frac{1}{\sum_{i=1}^{K} L_i} \sum_{i=1}^{K} \sum_{\tau=1}^{L_i}
\min\!\Big( \rho_{i,\tau} A_i,\;
\text{clip}(\rho_{i,\tau}, 1-\epsilon, 1+\epsilon)\, A_i \Big)
\right]\;,
\end{equation}

\noindent where $L_i$ is the number of valid tokens in $o_i$ and $\rho_{i,\tau}$ is the per-token importance ratio between the current policy $\pi_\theta$ and the behavior policy $\pi_{\theta_{\text{old}}}$:
\begin{equation}
\rho_{i,\tau} =
\frac{\pi_\theta(o_{i,\tau} \mid x, o_{i,<\tau})}
{\pi_{\theta_{\text{old}}}(o_{i,\tau} \mid x, o_{i,<\tau})}.
\end{equation}

\paragraph{Retrieval Reward.} Using retrieval metrics such as nDCG@$k$ directly as rewards yields sparse signal, since most rewrites leave the gold passage outside the metric cutoff. We adopt Rank-Incentive Reward Shaping (RIRS), also used by ConvSearch-R1~\cite{zhu2025convsearchr1}, which converts the rank of the gold passage into a dense piecewise reward: ranks 1 through 10 map linearly onto $[1, 2]$, ranks 11 through 100 map onto $[0, 1]$, and ranks beyond 100 receive zero. The reward is computed from the top-100 ranking produced by the final rewrite issued before \texttt{<stop>}, or from the last issued query if the step budget is exhausted without stopping. Although the agent observes only the top-$k$ passages returned at each step, reward computation uses the retriever's top-100 ranking. Trajectories that violate the action format receive zero reward.

\section{Experiments}

This section evaluates our multi-step query rewriting agent. We describe the experimental setup (\S\ref{sec:setup}), report in-domain results (\S\ref{sec:results}), analyze the learned trajectories (\S\ref{sec:analysis}), test generalization to unseen benchmarks and retrieval backends (\S\ref{sec:generalization}), and quantify the cost of sequential interaction (\S\ref{sec:overhead}).

\subsection{Experimental Setup}\label{sec:setup}

\paragraph{Datasets.}  We evaluate on three conversational search benchmarks. QReCC~\cite{anantha2021qrecc} covers query rewriting over a large collection, and we use its standard training (29,596 queries) and testing (8,209 queries) splits. TopiOCQA~\cite{adlakha-etal-2022-topiocqa} is built around topic switches, with longer conversations that span several topics, which makes it harder for query reformulation than QReCC~\cite{adlakha-etal-2022-topiocqa,mo2023convgqr}. We use its standard training set (45,450 queries) and development set (2,514 queries). Training uses half the queries for SFT, and the remaining half for RL. TREC CAsT~\cite{dalton2019cast,dalton2020cast} is used exclusively for zero-shot evaluation, i.e., models trained on QReCC or TopiOCQA are evaluated on CAsT without further training.

\paragraph{Retrievers.} Following prior work~\cite{yoon2025retpo,zhu2025convsearchr1}, we use BM25 ($k_1{=}0.90$, $b{=}0.40$ for TopioCQA; $k_1{=}0.82$, $b{=}0.68$ for QReCC) as our sparse retriever through Pyserini~\cite{DBLP:conf/sigir/LinMLYPN21}. For generalization experiments, we leverage ANCE~\cite{DBLP:conf/iclr/XiongXLTLBAO21} and Qwen3-Embedding-0.6B~\cite{2025_qwen3embedding} as dense retrievers through FAISS~\cite{johnson2019billion}. No retriever is trained or adapted in any of our experiments.

\paragraph{Baselines.} We compare against the raw conversational query and five trained rewriters: CHIQ-Fusion~\cite{mo2024chiq}, AdaQR~\cite{zhang2024adaqr}, ADACQR~\cite{lai2025adacqr}, RETPO~\cite{yoon2025retpo}, and ConvSearch-R1~\cite{zhu2025convsearchr1}, the strongest prior method and our primary point of comparison. All baseline numbers are taken from ConvSearch-R1's evaluation, which shares our evaluation splits, retriever, and metrics. We re-ran ConvSearch-R1's released inference code and weights and reproduced its reported scores. We also apply RM3~\cite{lavrenko2001relevance} to our agent's first rewriting step, contrasting a fixed single-round feedback mechanism with the agent's learned multi-step feedback. 

\paragraph{System Configurations.} We compare three configurations of our system, each trained with the same procedure and differing only in what the policy is allowed to do. The \textit{single-step, single-action} configuration emits one free-form rewrite and issues it once, reproducing the ConvSearch-R1 setting within our harness and isolating the contribution of the backbone from that of the method. The \textit{single-step, multi-action} configuration selects one action from the typed space and issues it once, isolating the value of typed rewriting from that of iteration. The \textit{multi-step, multi-action} configuration is the full agent, selecting a typed action at each step and conditioning on the passages retrieved so far. For each configuration, we report the untrained Qwen3-4B and Qwen3-32B backbones under the same prompt and harness, the checkpoint after supervised fine-tuning, and the checkpoint after reinforcement learning.

We use Qwen3-4B as the backbone for every trained configuration. The agent runs for at most $T = 5$ steps, and each search appends the top 5 passages to the context, each truncated to 128 tokens. At each step, the agent can generate up to 512 tokens, with a total trajectory budget of 8192 tokens covering both agent generations and retrieved passages. SFT trajectories are distilled from the agent running with Qwen3-32B as the backbone, with a separate SFT checkpoint per configuration, since the configurations differ in action space. Reinforcement learning uses GRPO with group size $K = 8$ and the rank-incentive reward that was described in Section~\ref{sec:training}. Additional details are shown on Appendix~\ref{app:1}.


\subsection{Results}\label{sec:results}

Table~\ref{tab:main} reports the main comparison. The full \textit{multi-step, multi-action} agent achieves the highest MRR and NDCG@3 among the compared methods, including both ConvSearch-R1 variants. The gains are largest on TopiOCQA, the benchmark with the most frequent topic shifts, and concentrate at the top of the ranking. The gains are therefore more evident in the precision-oriented metrics, indicating that successive rewrites primarily improve the top of the ranking rather than broaden its coverage.

The controlled configurations show that the gain comes from the combination of iteration and typed rewriting. On TopiOCQA, our \textit{single-step, single-action} setting, which mirrors the ConvSearch-R1 formulation, achieves performance consistent with that reported by ConvSearch-R1. Allowing the agent to choose among typed actions without allowing iteration instead reduces performance, showing that the typed action space alone is not beneficial. The full \textit{multi-step, multi-action} agent achieves the strongest results, demonstrating that typed actions become effective when the agent can sequence them in response to retrieved evidence. QReCC shows the same pattern, although with smaller (yet significant) gains, consistent with the dataset being more saturated, i.e., even relatively simple rewriting methods already achieve MRR above 50\%. 

It is also worth noting that the advantage of iteration is absent after supervised fine-tuning and emerges under reinforcement learning. The multi-step SFT checkpoint does not exceed the \textit{single-step, single-action} SFT checkpoint, i.e., the gap emerges only after RL optimizes against the retriever. Distillation from the 32B-parameter version of the model instills the action format, but the policy that exploits iteration is discovered through the retrieval reward rather than transferred from the distilled trajectories.

\begin{table}[t]
\centering
\caption{Retrieval results on TopiOCQA and QReCC over a sparse (BM25) backend. Baseline numbers come from ConvSearch-R1~\cite{zhu2025convsearchr1}. Bold indicates the best result per column, and $\dagger$~indicates a statistically significant improvement over ConvSearch-R1 re-run from its open weights (paired t-test over per-query scores, $p<0.05$).}
\label{tab:main}
\setlength{\tabcolsep}{3.2pt}
\begin{tabular}{l ccc ccc}
\toprule
& \multicolumn{3}{c}{\textbf{TopiOCQA}} & \multicolumn{3}{c}{\textbf{QReCC}} \\
\cmidrule(lr){2-4} \cmidrule(lr){5-7}
\textbf{Method} & MRR & nDCG@3 & R@10 & MRR & nDCG@3 & R@10 \\
\midrule
\rowcolor{gray!15}
\multicolumn{7}{l}{\textbf{Baselines}} \\
\cdashline{1-7}
\addlinespace[3pt]
Raw query                & 2.1  & 1.8  & 4.0  & 6.5  & 5.5  & 11.1 \\
Step-1 + RM3 & 18.0 & 15.6 & 36.6 & 37.7 & 34.4 & 59.0 \\
AdaQR                    & 20.3 & 18.0 & 37.1 & 50.6 & 48.0 & 69.6 \\
CHIQ-Fusion              & 25.6 & 23.5 & 44.7 & 54.3 & 51.9 & \textbf{78.5} \\
ADACQR                   & 28.3 & 26.5 & 48.9 & 55.1 & 52.5 & 76.5 \\
RETPO                    & 28.3 & 26.5 & 48.3 & 50.0 & 47.3 & 69.5 \\
ConvSearch-R1 (Llama3.2) & 37.8 & 36.2 & \textbf{59.6} & 55.9 & 54.3 & 77.2 \\
ConvSearch-R1 (Qwen2.5)  & 35.2 & 33.5 & 57.8 & 56.5 & 54.8 & 76.3 \\
\midrule
\rowcolor{gray!15}
\multicolumn{7}{l}{\textbf{Ours} (single step, single rewriting action)} \\
\cdashline{1-7}
\addlinespace[3pt]
Qwen3-32B & 12.2  & 11.5  & 21.8 & 18.4 & 17.4 & 26.3 \\
Qwen3-4B & 3.2  & 2.9  & 5.7  & 6.2   & 5.8   & 8.6   \\
\quad + SFT      & 20.7 & 18.8 & 38.5 & 42.7   & 40.4   & 61.9   \\
\quad + SFT + RL & 34.3 & 32.5 & 58.9 & 56.3   & 54.4   & 76.2 \\
\midrule
\rowcolor{gray!15}
\multicolumn{7}{l}{\textbf{Ours} (single step, multiple rewriting actions)} \\
\cdashline{1-7}
\addlinespace[3pt]
Qwen3-32B & 16.6  & 14.6  & 31.5 & 34.6 & 32.5 & 50.9 \\
Qwen3-4B & 3.4  & 2.9  & 6.8  & 7.0  & 6.5  & 10.8 \\
\quad + SFT      & 16.4 & 14.5 & 31.1 & 37.1 & 34.7 & 55.1 \\
\quad + SFT + RL & 24.6 & 22.5 & 44.6 & 55.7 & 53.7 & 75.0 \\
\midrule
\rowcolor{gray!15}
\multicolumn{7}{l}{\textbf{Ours} (multiple steps, multiple rewriting actions)} \\
\cdashline{1-7}
\addlinespace[3pt]
Qwen3-32B & 21.6  & 19.8  & 38.9 & 41.3 & 38.9 & 60.5 \\
Qwen3-4B & 9.7  & 8.5  & 18.4 & 22.5 & 20.9 & 34.2 \\
\quad + SFT      & 19.4 & 17.5 & 34.7 & 38.5 & 36.3 & 56.2 \\
\quad + SFT + RL & \textbf{41.4}$^\dagger$ & \textbf{40.6}$^\dagger$ & \textbf{59.6} & \textbf{57.2}$^\dagger$ & \textbf{55.8}$^\dagger$ & 75.8 \\
\bottomrule
\end{tabular}
\end{table}

\subsection{Analysis and Ablations}\label{sec:analysis}

\begin{figure}[t!]
    \centering
    \includegraphics[width=1\linewidth]{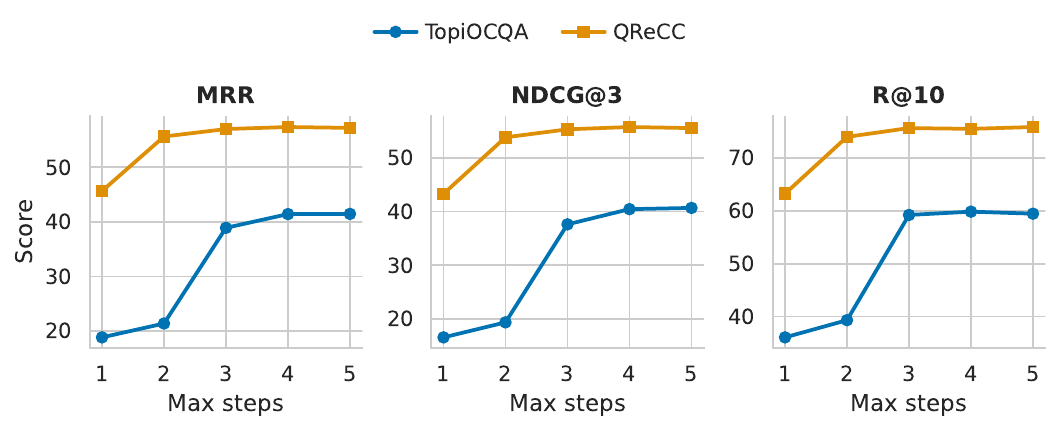}
    \caption{Retrieval quality on TopiOCQA and QReCC (BM25) as the 5-step agent is truncated at inference to a maximum of one to five steps. }
    \label{fig:budget}
\end{figure}

\begin{figure}[t!]
    \centering
    \includegraphics[width=1\linewidth]{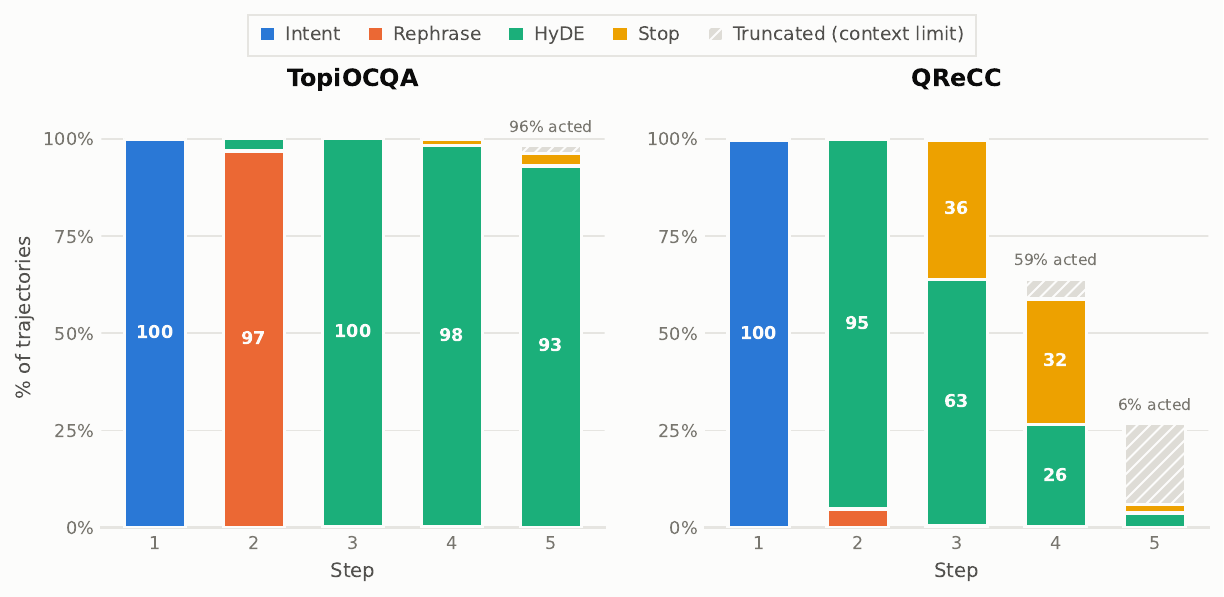}
    \caption{Action distribution at each step of the RL-trained policy on
TopiOCQA and QReCC (BM25). Bar heights are shares of all trajectories, and
numbers inside the bars give the share taking each action. A trajectory
stops acting once it emits \texttt{<stop>} or its context fills after
retrieved passages are appended (\emph{Truncated}). Label above
the bars give the fraction still acting when it falls below 100\%.}
    \label{fig:actions}
\end{figure}

\paragraph{Interaction Budget.}  Figure~\ref{fig:budget} truncates the 5-step agent at inference without retraining, so every point uses the same policy and differs only in how many reformulations it may issue. Retrieval quality rises steeply over the first three steps and then flattens. Truncated to a single step, the agent scores below the reinforcement-learned baselines in Table~\ref{tab:main}, suggesting that the improvement depends substantially on sequential interaction rather than on the backbone or typed rewriting actions alone.


\paragraph{Learned Action Order.} The policy converges to a consistent order that neither the prompt nor the reward enforces (Figure~\ref{fig:actions}). Intent resolution takes the first step of every trajectory on both datasets, and pseudo-document generation dominates the final one. SFT alone does not enforce this behavior. Rollouts from the SFT checkpoints average 1.7 search steps, and only 60\% open with intent resolution, the rest opening with a rephrase or a pseudo-document. Under RL these alternative openings disappear and episodes lengthen to 4.9 search steps on TopiOCQA and 2.9 on QReCC. Since the reward scores only the ranking of the final rewrite and encodes no preference over order or length, the ordering appears to emerge from retrieval-reward optimization rather than inherited from distillation. A single-step agent restricted to (ungrounded) pseudo-documents, trained with the same procedure, reaches 30.3 nDCG@3 on TopiOCQA, below both the free-form single-step rewriter (32.5) and the full agent (40.6). Pseudo-documents therefore help most when conditioned on passages retrieved by earlier steps.

The policy's transition to pseudo-document generation drives the performance gains observed in Figure~\ref{fig:budget}, marking the sharpest retrieval improvements in the episode (i.e., step three on TopiOCQA and step two on QReCC). A potential issue is that this operation might reduce to trivial re-ranking, i.e., the agent could copy one of the lower-ranked passages among the top-5 results and reissue it verbatim, which, since a passage submitted as a query should retrieve itself at rank one, would promote it without any genuine reformulation. To rule this out, for each pseudo-document we measure 8-gram containment, i.e. the fraction of its 8-word spans occurring verbatim in a passage, maximized over all passages in the agent's context, and the median was 0. The generation is nonetheless grounded in retrieved evidence, since unigram Jaccard between each pseudo-document and the passages (excluding stop-words) in its context gives a mean of 0.137, against 0.014 for passages drawn from an unrelated query's context, roughly an order of magnitude higher. Thus, the model incorporates vocabulary from the retrieved passages without simply copying them. Consistent with this, replacing the contextual passages at inference with those from an unrelated query reduces nDCG@3 on TopiOCQA from 40.6 to 25.3.


\subsection{Generalization Across Domains and Retrieval Backends}
\label{sec:generalization}

\begin{table}[t]
\centering
\caption{Zero-shot transfer retrieval results on TREC CAsT 2019 and 2020. Baseline numbers come from
ConvSearch-R1~\cite{zhu2025convsearchr1}.}
\label{tab:cast}
\setlength{\tabcolsep}{3.2pt}
\resizebox{\linewidth}{!}{%
\begin{tabular}{l cccc cccc}
\toprule
& \multicolumn{4}{c}{\textbf{CAsT 2019}} & \multicolumn{4}{c}{\textbf{CAsT 2020}} \\
\cmidrule(lr){2-5} \cmidrule(lr){6-9}
\textbf{Method} & MRR & NDCG@3 & R@10 & R@100 & MRR & NDCG@3 & R@10 & R@100 \\
\midrule
\rowcolor{gray!15}
\multicolumn{9}{l}{\textbf{Baselines}} \\
\cdashline{1-9}
\addlinespace[3pt]
Human Rewrite  & 74.0 & 46.1 & -- & 38.1 & 59.1 & 42.2 & -- & \underline{46.5} \\
LLM4CS  & 70.4 & 46.8 & 11.7 & -- & 58.6 & 41.5 & \textbf{19.3} & -- \\
CHIQ-Fusion  & 73.3 & 50.5 & 12.9 & -- & 54.0 & 38.2 & \textbf{19.3} & -- \\
ConvSearch-R1  & 76.6 & \underline{54.0} & \textbf{14.2} & 42.0 & \textbf{70.6} & \underline{42.6} & 15.6 & 43.8 \\
\midrule
\rowcolor{gray!15}
\multicolumn{9}{l}{\textbf{Ours} (multiple steps, multiple rewriting actions) -- Trained on QReCC} \\
\cdashline{1-9}
\addlinespace[3pt]
Qwen3-4B + SFT + RL & \underline{79.1}  & 51.9 & \textbf{14.2} & \textbf{48.4} & 66.4 & 38.8 & 16.2 & 45.2 \\
\midrule
\rowcolor{gray!15}
\multicolumn{9}{l}{\textbf{Ours} (multiple steps, multiple rewriting actions) -- Trained on TopiOCQA} \\
\cdashline{1-9}
\addlinespace[3pt]
Qwen3-4B + SFT + RL & \textbf{80.6}  & \textbf{54.7} & \underline{14.1} & \underline{48.0} & \underline{70.1} & \textbf{44.1} & \underline{17.4} & \textbf{47.6} \\
\bottomrule
\end{tabular}}
\end{table}

\begin{table}[t]
\centering
\caption{Retrieval results of the \textit{multi step, multi action} agent on TopiOCQA and QReCC when changing search backend during inference.}
\label{tab:retrievers}
\setlength{\tabcolsep}{3.2pt}
\begin{tabular}{l ccc ccc}
\toprule
& \multicolumn{3}{c}{\textbf{TopiOCQA}} & \multicolumn{3}{c}{\textbf{QReCC}} \\
\cmidrule(lr){2-4} \cmidrule(lr){5-7}
\textbf{Inference Retriever} & MRR & NDCG@3 & R@10 & MRR & NDCG@3 & R@10 \\
\midrule
BM25  & 41.4 & 40.6 & 59.6 & 57.2 & 55.8 & 75.8 \\
ANCE  & 44.0 & 43.4 & 64.4 & 47.0 & 45.0 & 66.6 \\
Qwen3-Embedding-0.6B  & 51.0 & 50.6 & 72.4 & 54.1 & 52.1 & 75.0 \\
\bottomrule
\end{tabular}
\end{table}

We next examine whether the learned multi-step rewriting policy generalizes beyond the retrieval setting used for training. We consider two forms of transfer: zero-shot evaluation on an unseen conversational search benchmark, and changing the retrieval backend at inference time without retraining the policy.

Table~\ref{tab:cast} evaluates the \textit{multi-step, multi-action} agent on TREC CAsT 2019 and 2020. Both policies transfer to CAsT without further training, with the TopiOCQA-trained policy being consistently the stronger of the two. A plausible explanation is closer alignment between TopiOCQA and CAsT, since TopiOCQA is built around topic shifts within a conversation, a property that CAsT conversations also exhibit, and its Wikipedia collection partially overlaps with the TREC CAR portion of the CAsT corpus. The TopiOCQA-trained model exceeds ConvSearch-R1 on NDCG@3 in both CAsT years, suggesting that the learned interaction between typed rewriting and retrieval feedback is not specific to the conversational distribution of the training data.


Table~\ref{tab:retrievers} evaluates whether a policy trained against BM25 transfers to dense retrievers at inference time, without additional fine-tuning. On TopiOCQA, both dense backends improve over BM25 on every metric, with Qwen3-Embedding yielding the largest gains, indicating that the learned rewriting behavior is not tied to the retriever it was trained with. On QReCC, neither dense backend surpasses BM25, although the gap narrows considerably with the stronger embedding model, consistent with prior work reporting BM25 ahead of dense retrievers on this benchmark~\cite{zhu2025convsearchr1}. A plausible factor is that QReCC answers share substantial word-level overlap with their supporting passages~\cite{zhang2024adaqr}, which favours lexical matching once the rewrite restores the missing context. Overall, the policy transfers across retriever paradigms, while absolute performance still depends on how well the retriever suits the dataset.

\subsection{Cost of Sequential Interaction}
\label{sec:overhead}

\begin{figure}[t!]
    \centering
    \includegraphics[width=1\linewidth]{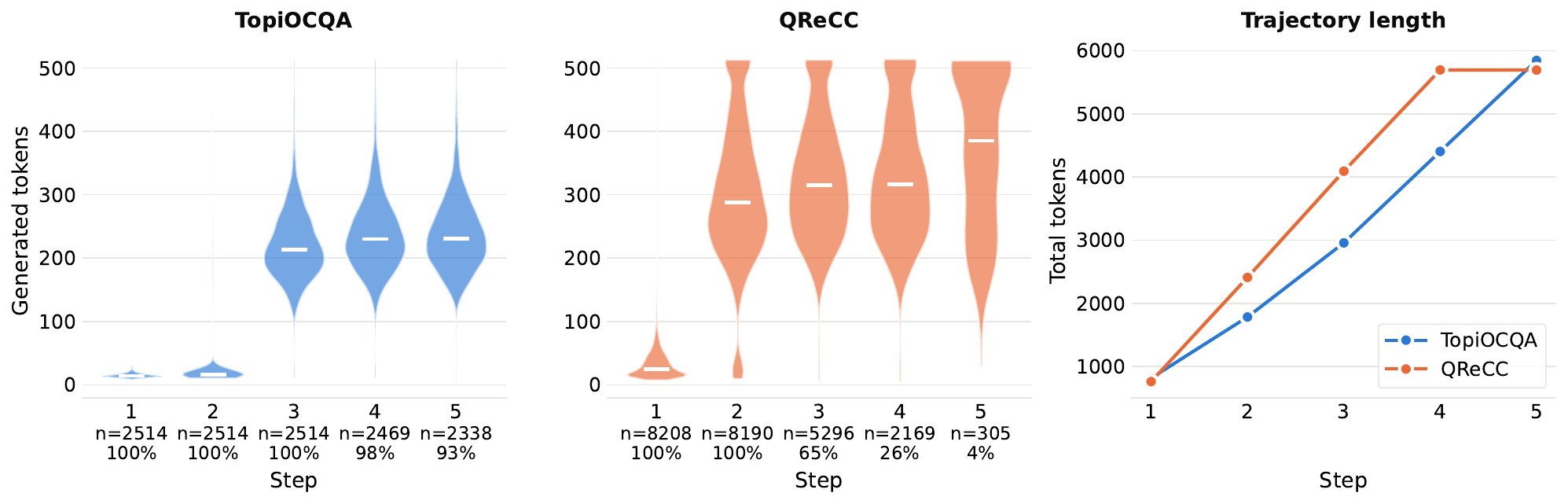}
    \caption{\textbf{(left, middle)} Distribution of tokens generated per step, over trajectories that emit a rewrite at that step (i.e., \texttt{<stop>} not considered). \textbf{(right)} Mean trajectory length, i.e. prompt, actions and retrieved passages accumulated up to that step. The QReCC curve flattens at step 5 because longer trajectories have already stopped.}
    \label{fig:overhead}
\end{figure}

We quantify the inference cost of sequential interaction in terms of generated tokens. We compare against ConvSearch-R1~\cite{zhu2025convsearchr1} run from its released weights. As a single-step rewriter, ConvSearch-R1 generates approximately 400 tokens per example on both datasets. Our individual steps are shorter, e.g., intent resolution and lexical reformulation typically produce only a few dozen tokens, while average pseudo-document synthesis remains below the length of a single ConvSearch-R1 generation (Figure~\ref{fig:overhead} left and middle). Sequential interaction nevertheless increases total generation cost because each additional step is generated over an expanding context containing previously retrieved passages. The resulting cost is dataset-dependent. On TopiOCQA, 93\% of trajectories issue a fifth rewrite, making the later interactions the main source of additional generation cost, whereas on QReCC only 4\% issue a fifth rewrite. Thus, although individual rewriting actions are inexpensive, their cumulative cost depends on how long the learned policy continues interacting with the corpus.
\section{Conclusions and Future Work}

In this paper we studied conversational query rewriting as a sequential
retrieval process driven by feedback from intermediate search results. We
trained a multi-step agent with reinforcement learning and no human rewrite
annotations. Rewarded solely on retrieval quality, the policy converges to a
consistent action order: it first resolves conversational intent,
optionally refines the lexical formulation, and then generates
pseudo-documents grounded in the passages retrieved by earlier steps. This
formulation achieves the strongest ranking precision among the compared
methods on TopiOCQA and QReCC, transfers zero-shot to TREC CAsT, and remains effective under dense retrievers.

Several directions extend from this sequential framework. While we leveraged the RIRS reward directly for better comparison with prior studies, incorporating step penalties or latency costs into the reward objective could yield cost-aware policies that adaptively terminate early when low conversational ambiguity permits. Also, the agent could issue multiple reformulations in parallel at each step and be rewarded on their fused ranking, encouraging complementary rewrites that broaden recall while shortening the sequential loop.


%

\appendix
\renewcommand{\theHsection}{appendix.\arabic{section}} 
\section{Reproducibility Details}\label{app:1}

The system prompt of the \textit{multi-step, multi-action} agent is shown below. RL uses AReaL~\cite{DBLP:conf/nips/FuGSZMHXWMWYYW25} with full fine-tuning, Adam at learning rate 5e-6 with a constant schedule, 16 prompts per step with 8 rollouts each, symmetric clipping $\epsilon$ = 0.4, and no KL penalty. Training runs one epoch over the RL split (1,420 steps on TopiOCQA, 930 on QReCC), and we report the final checkpoint. SFT uses 7,175 (TopiOCQA) and 7,682 (QReCC) Qwen3-32B trajectories, trained for one epoch at learning rate 2e-5. All generation samples at temperature 1.0 with Qwen3's thinking mode enabled, although the RL policies stop emitting reasoning blocks. Training used four NVIDIA L40S GPUs.

\begin{promptbox}[title={System prompt: multi-step, multi-action rewriting}]
\scriptsize
\setlength{\parskip}{2pt}

You are a query rewriting assistant for conversational search. You help resolve ambiguous, context-dependent conversational queries into effective standalone search queries.

You will receive a conversational query line in the following format:
\begin{itemize}[leftmargin=1.1em, nosep]
  \item The string contains the \textbf{most recent user query first}, followed by zero or more \lit{[SEP]}-separated snippets representing \textbf{dialogue history} (earlier turns, which may include prior questions and/or short answers or facts).
  \item Format: \lit{CURRENT\_QUERY [SEP] HISTORY\_ITEM\_1 [SEP] HISTORY\_ITEM\_2 [SEP] ...}
  \item History is ordered from most recent to oldest as you move right.
\end{itemize}

Your job is to rewrite the \textbf{current (leftmost) query only}, using the dialogue history purely for disambiguation and reference resolution.

\phead{\# Workflow}
At each step you will:
\begin{enumerate}[leftmargin=1.4em, nosep]
  \item Think about what the user is looking for and what gaps remain in the current results.
  \item Choose exactly one action from the four below.
\end{enumerate}

\phead{\# Actions}
You have four actions. Each submits a single query to the search engine and returns results.
\begin{enumerate}[leftmargin=1.4em, nosep, itemsep=3pt]
  \item \textbf{Intent query} --- a focused standalone query capturing the user's underlying information need. Resolve all coreferences and ellipsis.\\
        \thinktag{}\, \ptag{intent}\textcolor{gray}{\itshape focused standalone query}\ptag{/intent}
  \item \textbf{Rephrase query} --- the same information need expressed with alternative terminology, synonyms, or different phrasings. Think about how this topic might appear in encyclopedic or reference text.\\
        \thinktag{}\, \ptag{rephrase}\textcolor{gray}{\itshape alternative phrasing of the query}\ptag{/rephrase}
  \item \textbf{HyDE query} --- a short hypothetical passage that would correctly answer the question, written using your own knowledge. This introduces answer-side vocabulary that helps the retriever match relevant documents.\\
        \thinktag{}\, \ptag{hyde}\textcolor{gray}{\itshape short hypothetical answer passage}\ptag{/hyde}
  \item \textbf{Stop} --- submit the current retrieved results if they look relevant, or if further rewriting is unlikely to help.\\
        \thinktag{}\, \ptag{stop}\ptag{/stop}
\end{enumerate}

\phead{\# Guidelines}
\begin{itemize}[leftmargin=1.1em, nosep]
  \item Resolve coreferences: replace pronouns (``he'', ``it'', ``they'') with the actual entities from conversation history.
  \item Resolve ellipsis: fill in omitted information from context.
  \item Handle topic shifts: if the conversation has shifted topics, focus on the current topic.
  \item \textbf{Gap-bridging}: after seeing retrieved passages, identify what's missing and choose the action type most likely to bridge the gap.
  \item \textbf{Adopt corpus vocabulary}: if the retrieved passages use specific terminology (names, dates, technical terms), incorporate those exact terms.
  \item Stop early if results look relevant. Don't rewrite unnecessarily.
  \item You MUST output exactly one action per turn.
\end{itemize}
\end{promptbox}

\bibliographystyle{splncs04}
\bibliography{bib}

\end{document}